# Computing Approximate Nash Equilibria and Robust Best-Responses Using Sampling

**Marc Ponsen** M.PONSEN@MAASTRICHTUNIVERSITY.NL
**Steven de Jong** STEVEN.DEJONG@MAASTRICHTUNIVERSITY.NL
*Department of Knowledge Engineering*
*Maastricht University, The Netherlands*

**Marc Lanctot** LANCTOT@UALBERTA.CA
*Department of Computer Science*
*University of Alberta, Canada*

## Abstract

This article discusses two contributions to decision-making in complex partially observable stochastic games. First, we apply two state-of-the-art search techniques that use Monte-Carlo sampling to the task of approximating a Nash-Equilibrium (NE) in such games, namely Monte-Carlo Tree Search (MCTS) and Monte-Carlo Counterfactual Regret Minimization (MCCFR). MCTS has been proven to approximate a NE in perfect-information games. We show that the algorithm quickly finds a reasonably strong strategy (but not a NE) in a complex *imperfect* information game, i.e. Poker. MCCFR on the other hand has theoretical NE convergence guarantees in such a game. We apply MCCFR for the first time in Poker. Based on our experiments, we may conclude that MCTS is a valid approach if one wants to learn reasonably strong strategies fast, whereas MCCFR is the better choice if the quality of the strategy is most important.

Our second contribution relates to the observation that a NE is not a best response against players that are not playing a NE. We present Monte-Carlo Restricted Nash Response (MCRNR), a sample-based algorithm for the computation of restricted Nash strategies. These are robust best-response strategies that (1) exploit non-NE opponents more than playing a NE and (2) are not (overly) exploitable by other strategies. We combine the advantages of two state-of-the-art algorithms, i.e. MCCFR and Restricted Nash Response (RNR). MCRNR samples only relevant parts of the game tree. We show that MCRNR learns quicker than standard RNR in smaller games. Also we show in Poker that MCRNR learns robust best-response strategies fast, and that these strategies exploit opponents more than playing a NE does.

## 1. Introduction

This article investigates decision-making in strategic, complex multi-player games. As our most complex test-bed, we use the game of two-player Limit Texas Hold'em Poker (henceforth abbreviated as 'Poker' or 'full-scale Poker'). In this introduction, we will first briefly outline why research in such games is relevant. Then, we discuss the complexity factors involved in games. Finally, we outline our approach and contributions.

### 1.1 Relevance of Games-Related Research

Games have attracted scientific attention for years now; the importance of research in the area of game theory became apparent during the Second World War (Osborne & Rubinstein, 1994). Nowadays, examples of serious games can be found in many real-life endeavors, such as economics (e.g., buyers and sellers in the stock market have the goal to maximize profit) or politics (e.g., politicians have the goal to collect sufficient political support for some cause). Games that serve as entertainment, such as puzzles, board-, sports- or modern video-games, are often abstracted, simpler variants of serious games. An example is the card game of Poker. The objective in Poker is not very different from the objective of investors in the stock market. Players may invest (or risk) money and speculate on future events that may or may not yield profit. Because strategies in abstract games (such as Poker) can be more easily and rapidly evaluated than strategies in real-life endeavors (such as acting in the stock market), abstract games are the perfect tool for assessing and improving the strategic decision-making abilities of humans as well as computers. For this reason, various complex multi-player games have attracted a great deal of attention from the artificial intelligence (AI) community (Schaeffer, 2001).

### 1.2 Complexity Factors in Games

Games are characterized by several complexity factors. We briefly mention some factors that are relevant for the work presented in this article.

- *Number of players*. Multi-player games are generally assumed to be more complex than single-player games. Within the class of multi-player games, fully competitive games, also known as zero-sum games, are those games where all players have conflicting goals and therefore deliberately try to minimize the payoff of others.

- *Size of the state space*. The size of the state space (the number of different situations the game may be in) varies from game to game, depending on the number of legal situations and the number of players. Large state spaces produce more complex games because of the computational requirements for traversing the entire state space.

- *Uncertainty*. Stochastic games, as opposed to deterministic games, are more complex because there is uncertainty about the effects of actions, or occurrences of (future) events, for instance because die rolls are involved.

- *Imperfect information*. Parts of the game state may be hidden to the players, e.g., opponent cards in any card game, or the probability of a certain chance outcome. This is also known as partial observability.

In the remainder of this article, we deal with partially observable stochastic games (Fudenberg & Tirole, 1991), using a full-scale Poker game as our most complex test-bed. The game is a multi-player, competitive, partially observable stochastic game. It is a daunting game for both human and AI players to master.

### 1.3 Our Contributions

We investigate two different sampling-based approaches for decision-making, namely (1) a classical game-theoretic approach and (2) a best-response approach.

For our first approach, we apply current state-of-the-art algorithms to the task of computing an approximated Nash-Equilibrium strategy (NES) in the game of Poker. In a two-player, zero-sum game, the expected value of a NES is a constant, regardless of the opponent strategy or the specific NES. In fair games (i.e. players have equal chance of winning), such strategies cannot lose in expectation and may win in the long run. For complex games such as Poker, a NES can only be computed by introducing abstractions. Also, sampling algorithms may be used to (relatively) quickly compute an approximate NES. We use both abstractions as well as sampling in this work. We will look at two families of algorithms, both of which rely on Monte-Carlo sampling, namely Monte-Carlo Tree Search (MCTS), including Upper Confidence Bounds applied to Trees (Kocsis & Szepesvári, 2006; Chaslot, Saito, Bouzy, Uiterwijk, & van den Herik, 2006; Coulom, 2006), and a regret-minimizing algorithm, called Monte-Carlo Counterfactual Regret Minimization (MCCFR) (Lanctot, Waugh, Zinkevich, & Bowling, 2009). We are the first to offer a comparison between these two algorithms in full-scale Poker.[1]

For our second approach, we begin with the observation that a NES is not necessarily most profitable against any strategy other than a NES. Above all, it is a *safe* strategy. If we have information on the strategy of opponent players, we can adapt our strategy based on this, i.e., by learning so-called best-response strategies. Rather than playing the safe NES (i.e., play not to lose), we want to learn tailored counter-strategies based on an opponent model (i.e., play to win). For learning a good compromise between best response and equilibrium, we combine the general technique of Restricted Nash Response (RNR) (Johanson, Zinkevich, & Bowling, 2008) with the Monte-Carlo Counterfactual Regret Minimization (MCCFR) algorithm, leading to a new algorithm, named Monte-Carlo Restricted Nash Response (MCRNR).

### 1.4 Structure of this Article

The remainder of this article is structured as follows. Section 2 provides a brief overview of the background knowledge required for this article, i.e., game-theoretic concepts focussed on extensive-form games, and a discussion on the games used in the article. Section 3 contains our work on a comparison of Monte-Carlo Tree Search (MCTS) and Monte-Carlo Counterfactual Regret Minimization (MCCFR) in full-scale Poker. Section 4 introduces Monte-Carlo Restricted Nash Response (MCRNR) and describes a set of experiments in smaller games as well as Poker. Finally, in Section 5, we conclude the article.

## 2. Background

In the current section we provide some background information. In Section 2.1 we discuss game theory (GT) fundamentals, focussing on extensive-form games. We can represent partially observable stochastic games by means of such games (Fudenberg & Tirole, 1991). Our main test domain, two-player limit Texas Hold'em Poker, is introduced in Section 2.2, along with a number of smaller games we also use to validate our new algorithm experimentally.

### 2.1 Game Theory and Extensive Form Games

Game theory (GT) studies strategic decision-making in games with two or more players. The basic assumptions that underlie the theory are that players are *rational*, i.e. they are self-interested and able to optimally maximize their payoff, and that they take into account their knowledge or

1. We note that we do not aim to join the 'arms race' to compute the closest Nash-Equilibrium (NE) approximation for full-scale Poker. The aim of our contribution is a comparison of two recent promising algorithms. Poker is the test-bed we chose to use because it is the most complex partially observable stochastic game both algorithms have been applied to thus far, and because there exist reasonably strong benchmarks to test against.

expectations of other decision-makers' behavior, i.e. they reason strategically (Fudenberg & Tirole, 1991; Osborne & Rubinstein, 1994). The field originated in economics to analyze behaviour in non-cooperative settings, and was firmly established by von Neumann and Morgenstern (1944). Nash (1951) introduced what is now known as the Nash-Equilibrium (NE). In the current section we briefly discuss the fundamentals of GT and extensive-form games.

#### 2.1.1 GAMES

Games are descriptions of strategic interaction between players. They specify a set of available actions to players and a payoff for each combination of actions. Game-theoretic tools can be used to formulate solutions for classes of games and examine their properties. Typically, a distinction is made between two types of game representations, namely normal-form games and extensive-form games. A normal-form game is usually represented by a matrix which shows the players, actions, and payoffs. In normal-form games it is presumed that all players act simultaneously (i.e. not having any information on the action choice of opponents). The second representation, the extensive-form game, describes how games are played over time. Previous action sequences are stored in a so-called game-tree, and as such, information about the choices of other players can be observed. In this article, we focus on extensive-form games.

#### 2.1.2 EXTENSIVE-FORM GAMES

An extensive-form game is a general model of sequential decision-making with imperfect information. As with perfect-information games (such as Chess or Checkers), extensive-form games consist primarily of a game tree where nodes represent states of the game. Each non-terminal node has an associated player (possibly chance) that makes the decision at that node, and each terminal node (leaf) has associated utilities for the players. Additionally, game states are partitioned into information sets $\mathcal{I}_i$. A player $i$ cannot distinguish between states in the same information set. The player, therefore, must choose actions with the same policy at each state in the same information set.

A **strategy** of player $i$, $\sigma_i$, is a function that assigns a probability distribution over $A(I_i)$ to each $I_i \in \mathcal{I}_i$, where $I_i$ is an information set belonging to $i$, and $A(I_i)$ is the set of actions that can be chosen at that information set. We denote $\Sigma_i$ as the set of all strategies for player $i$, and $\sigma_i \in \Sigma_i$ as the player's current strategy. A **strategy profile**, $\sigma$, consists of a strategy for each player, $\sigma_1, \ldots, \sigma_n$. We let $\sigma_{-i}$ refer to the strategies in $\sigma$ excluding $\sigma_i$.

Valid sequences of actions in the game are called **histories**, denoted $h \in H$. A history is a **terminal history**, $h \in Z$ where $Z \subset H$, if its sequences of actions lead from root to leaf. A **prefix history** $h \sqsubseteq h'$ is one where $h'$ can be obtained by taking a valid sequence of actions from $h$. Given $h$, the current player to act is denoted $P(h)$. Each information set contains one or more valid histories. The standard assumption is perfect recall: information sets are defined by the information that was revealed to each player over the course of a history, assuming infallible memory.

Let $\pi^\sigma(h)$ be the probability of history $h$ occurring if all players choose actions according to $\sigma$. We can decompose $\pi^\sigma(h)$ into each player's contribution to this probability. Here, $\pi_i^\sigma(h)$ is the contribution to this probability from player $i$ when playing according to $\sigma$. Let $\pi_{-i}^\sigma(h)$ be the product of all players' contribution (including chance) except that of player $i$. Finally, let $\pi^\sigma(h, z) = \pi^\sigma(z)/\pi^\sigma(h)$ if $h \sqsubseteq z$, and zero otherwise. Let $\pi_i^\sigma(h, z)$ and $\pi_{-i}^\sigma(h, z)$ be defined similarly. Using this notation, we can define the expected payoff for player $i$ as $u_i(\sigma) = \sum_{h \in Z} u_i(h)\pi^\sigma(h)$.

Given a strategy profile, $\sigma$, we define a player's **best response** as a strategy that maximizes their expected payoff assuming all other players play according to $\sigma$. The **best-response value** for player $i$ is the value of that strategy, $b_i(\sigma_{-i}) = \max_{\sigma'_i \in \Sigma_i} u_i(\sigma'_i, \sigma_{-i})$. An $\epsilon$-Nash-Equilibrium (NE) is an approximation of a best response against itself. Formally, an $\epsilon$-Nash-Equilibrium (NE) it is a strategy profile $\sigma$ that satisfies:

$$\forall i \in N \quad u_i(\sigma) + \epsilon \geq \max_{\sigma'_i \in \Sigma_i} u_i(\sigma'_i, \sigma_{-i}) \tag{1}$$

If $\epsilon = 0$ then $\sigma$ is a **Nash-Equilibrium (NE)**: no player has any incentive to deviate as they are all playing best responses. If a game is two-player and zero-sum, we can use **exploitability** as a metric for determining how close $\sigma$ is to an equilibrium, $\epsilon_\sigma = b_1(\sigma_2) + b_2(\sigma_1)$.

It is well-known that in two-player, zero-sum games NE strategies are *interchangeable*. That is, if $(\sigma_1, \sigma_2)$ and $(\sigma'_1, \sigma'_2)$ are different equilibrium profiles with $\sigma_i \neq \sigma'_i$, then $(\sigma_1, \sigma'_2)$ and $(\sigma'_1, \sigma_2)$ are also both NE profiles. This property makes equilibrium strategies for this class of games desireable since worst-case guarantees are preserved regardless of how the opponent plays. The property is easily extended to the case where $\epsilon > 0$, therefore playing an $\epsilon-$NE strategy will guarantee that a player is only exploitable by $\epsilon$. For more details, we refer the reader to the work of Fudenberg and Tirole (1991) as well as Osborne and Rubinstein (1994).

Throughout this article, we will refer to a player that plays the NES as a *rational* player. A player that plays rationally also assumes rationality on the part of its opponents. Experiments have shown that assuming rationality is generally not correct (e.g. for experiments in Poker see Billings, Burch, Davidson, Holte, Schaeffer, Schauenberg, & Szafron, 2003); even experienced human players in complex games at best play an approximated rational strategy, and frequently play dominated actions (i.e., actions that should never have been chosen at all). Moreover, for complex games such as Chess or Poker, even AI algorithms running on modern computers with a great deal of processor speed and memory can not (yet) cope with the immense complexity required to compute a NES. Thus, they are forced to either abstract the full game or do selective sampling to compute only an approximated NE.

### 2.2 Test Domains

In the current section, we introduce the test domains used throughout this article. In particular we describe the game of Poker, as well as some smaller games, some of which are similar to Poker. Poker is a card game played between at least two players. In a nutshell, the objective of the game is to win (money) by either having the best card combination at the end of the game (i.e. the showdown), or by being the only active player. The game includes several betting rounds wherein players are allowed to invest money. Players can remain active by at least matching the largest investment made by any of the players, or they can choose to fold (stop investing money and forfeit the game). In the case that only one active player remains, i.e. all other players chose to fold, the active player automatically wins the game. The winner receives the money invested by all the players. There exist many variants of the game. We will now specifically describe the ones used in this article.

**Kuhn Poker** is a two-player simple Poker game. There are only three cards (J - Jack, Q - Queen, and K - King). There are two actions, *bet* and *pass*. In the event of a showdown, the player with the higher card wins the pot (the King is highest and the Jack is lowest). After the deal, the first player has the opportunity to bet or pass. If the first player bets in round one, then in round two the second player can either bet (and go to showdown) or pass (and forfeit the pot). If the first player passes in round one, then in round two the second player can bet or pass. A bet leads to a third action for the first player, namely bet (and go to showdown) or pass (and forfeit the pot), whereas with a pass the game immediately proceeds to a showdown. Figure 1 shows the game tree with the first players' value for each outcome. The dominated actions have been removed from this tree. They include actions such as betting with a Queen as a first action, or passing a King for the second player. There are in total seven possible dominated actions to be made. The Nash-Equilibria of the game can be summarized by three parameters $(\alpha, \beta, \gamma)$ (each in $[0, 1]$). Kuhn determined that the set of equilibrium strategies for the first player has the form $(\alpha, \beta, \gamma) = (\gamma/3, (1+\gamma)/3, \gamma)$. Thus, there is a continuum of Nash-Equilibrium strategies governed by a single parameter $\gamma$. There is only one NES for the second player, namely $\eta = 1/3$ and $\xi = 1/3$. If either player plays a NES, then the first player expects to lose at a rate of $-1/18$ bets (i.e., size of ante) per hand. Kuhn Poker therefore is not a balanced or fair game, since the first player is expected to lose if both players play the rational strategy.

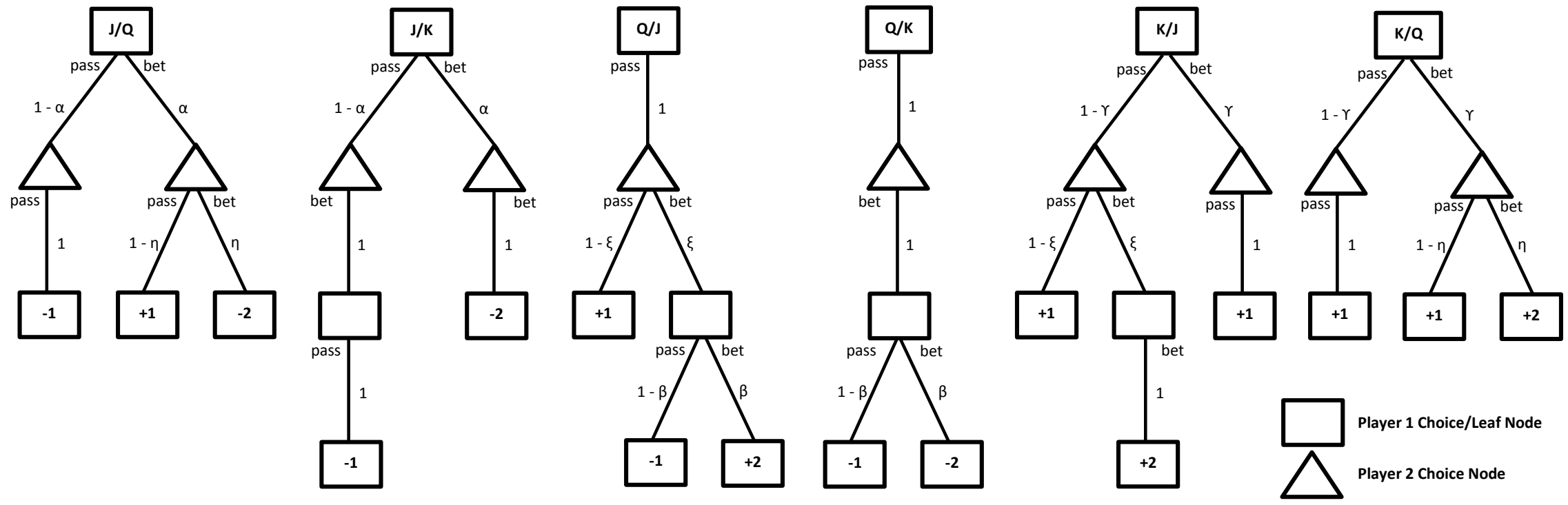

Figure 1: Kuhn Poker game tree (taken from the work of Hoehn, Southey, & Holte, 2005).

**One-Card Poker** (abbreviated OCP($N$)) (Gordon, 2005) is a generalization of Kuhn Poker. The deck contains $N$ cards; Kuhn Poker corresponds with $N = 3$. Each player must ante a single chip, has one more chip to bet with, and is dealt one card.

**Goofspiel** (abbreviated Goof($N$)) is a bidding card game where players have a hand of cards numbered 1 to $N$, and take turns secretly bidding on the top point-valued card in a point card stack, using cards in their hands (Ross, 1971). Our version is less informational: players only find out the result of each bid and not which cards were used to bid, and the player with the highest total points wins. We also use a fixed point card stack that is strictly decreasing, e.g. $(N, N-1, \dots, 1)$.

**Bluff(1,1,N)** also known as Liar's Dice and Perudo[2], is a dice-bidding game. In our version, each player rolls a single $N$-sided die and looks at their die without showing it to their opponent. Then players, alternately, either increase the current bid on the outcome of all die rolls in play, or call the other player's bluff (claim that the bid does not hold). The highest value on the face of a die is

2. See e.g. http://www.perudo.com/perudo-history.html.

wild and can count as any face value. When a player calls bluff, they win if the opponent's bid is incorrect, otherwise they lose.

**Texas Hold'em Poker** (Sklansky, 2005) is the most complex game under investigation here. The game includes 4 betting rounds, respectively called the preflop, flop, turn and river phase. During the first betting round, all players are dealt two private cards (i.e. only known to the specific player) out of a full deck consisting of 52 cards. To encourage betting, two players are obliged to invest a small amount the first round (the so-called small- and big-blind). One by one, the players can decide whether or not they want to participate in this game. If they indeed want to participate, they have to invest at least the current bet (i.e., the big-blind in the beginning of a betting round). This is known as *calling*. Players may also decide to *raise* the bet. If they do not wish to participate, players *fold*, resulting in loss of money they may have bet thus far. In the situation of no outstanding bet, players may choose to *check* (i.e., not increase the stakes) or *bet* more money. The size of bets and raises can either be predetermined (i.e., *Limit* Poker, as used in this paper), no larger than the size of the pot (i.e., *Pot-Limit* Poker) or unrestrained (i.e., *No-Limit* Poker). During the remaining three betting phases, the same procedure is followed. In every phase, community cards appear on the table (three cards in the flop phase, and one card in the other phases). These cards apply to all the players and are used to determine the card combinations (e.g., a pair or three-of-a-kind may be formed from the player's private cards and the community cards). During the showdown, if two or more players are still active, cards are compared, thus ending the game.

**Princess and Monster** (abbreviated PAM($R$, $C$, $H$)) (Isaacs, 1965) is not a Poker game; rather, it is a variation of the pursuit-evasion game on a graph, neither player ever knowing the location of the other nor discovering their moves (pursuit in a "dark room"). In our experiments we use random starting positions on a 4-connected grid graph with $R$ rows and $C$ columns. Players take turns alternately moving to an adjacent location. The game ends when the monster moves to the same location as the princess, or $H$ moves have been taken in total with no capture. The payoff to the evader is the number of steps uncaptured.

In the next section, we will use Kuhn Poker (or OCP(3)) and two-player Limit Texas Hold'em Poker. In Section 4, we use all games mentioned above, except Kuhn Poker; instead of Kuhn Poker, we use a larger game, OCP(500).

## 3. Computing Approximated Nash-Equilibrium Strategies

Our contribution in this section is to evaluate current and promising state-of-the-art search methods for computing (approximated) Nash-Equilibrium strategies in complex domains, and more specifically in Poker. Given the size of the game tree, such methods should (1) incorporate appropriate abstractions, and (2) be capable of analysing small subsets of the full game (i.e., do sampling).

We will look at two families of sampling algorithms, namely an Upper Confidence Bounds applied to Trees (UCT) based algorithm, called Monte-Carlo Tree Search (MCTS) (Kocsis & Szepesvári, 2006; Chaslot et al., 2006; Coulom, 2006), and a regret minimizing algorithm called Monte-Carlo Counterfactual Regret Minimization (MCCFR) (Lanctot et al., 2009). MCTS has achieved tremendous success in perfect information games, and in particular in the game of Go (Lee, Wang, Chaslot, Hoock, Rimmel, Teytaud, Tsai, Hsu, & Hong, 2010). In games with imper-fect information, it has no convergence guarantees for finding Nash-Equilibria.[3] However, it has been reported that it may nonetheless produce strong players in games with imperfect information (Sturtevant, 2008). We will empirically evaluate the merits of MCTS in the imperfect information game of Poker. MCCFR does have theoretical guarantees for convergence to a Nash-Equilibrium (NE). It has only been applied to smaller games thus far; we are the first to evaluate it in the complex domain of Poker.

This section is structured as follows. In Sections 3.1 and 3.2, we discuss existing work for computing Nash-Equilibrium strategies in large extensive games and provide details on the two sampling algorithms. Next, we will analyze the algorithms empirically, both in the domain of Kuhn Poker and also in the larger domain of two-player Limit Texas Hold'em Poker (Section 3.3).

### 3.1 Non-Sampling Algorithms for Computing Approximate Nash Equilibria

In the current subsection, we give an overview of existing non-sampling techniques for computing NE approximations in extensive form games. A large body of such work exists in many domains, but here we will focus specifically on work in the domain of Poker.

The conventional method for solving extensive form games (such as Poker), is to convert them into a linear program, which is then solved by a linear programming solver. Billings et al. (2003) solved an abstraction of the full game of two-player Poker using sequence-form linear programming. Their abstractions were three-fold. First, they learned models in separation for the different phases of the game. Basically, phases are considered independent and are solved in isolation. However, they state that previous phases contain important contextual information that is critical for making appropriate decisions. The probability distribution over cards for players strongly depends on the path that led to that decision point. Therefore, they provide input (i.e., contextual information such as pot size and the number of bets in the game) to the models learned for later phases. Second, they shorten the game tree by allowing only three (instead of the regular four) bet actions per phase. Third, they apply bucketing to the cards. The strategic strength of cards (i.e., private cards, possibly in combination with board cards) can be reflected by a numeric value. A higher value reflects stronger cards. Billings (2006) computed the so-called *Effective Hand Strength (EHS)*, a value in the range from 0 to 1, that combines the *Hand Strength* (i.e., current winning probability against active opponents) and *Hand Potential* (i.e., the probability that currently losing/winning cards end up winning/losing with the appearance of new board cards). The work of Billings (2006), in Sections 2.5.2.1 to 2.5.2.3, provides a more detailed discussion. Then they divide all values in buckets. For example, in a 10-bucket discretization (equal width), which we employ in this paper, EHS values in the range of 0 to 0.1 are grouped together in bucket 1, 0.1 to 0.2 in bucket 2, and so forth. A coarser view implies more information loss. The solution to the linear program induces a distribution over actions at each information set, which corresponds to a mixed behavioral strategy. A Poker-playing program can then sample actions from this mixed strategy. The resulting Poker-playing program was competitive with human experts.

The Counterfactual Regret Minimization (CFR) algorithm (Zinkevich, Johanson, Bowling, & Piccione, 2008) may be used to compute an approximated NE in richer abstractions because it requires less computational resources. In Poker, Zinkevich et al. (2008) only applied abstraction on the cards and were capable to learn strategies using CFR that were strong enough to defeat

3. For example, Shafiei, Sturtevant, and Schaeffer (2009) provide an analysis in simultaneous imperfect-information games, indicating that UCT finds suboptimal solutions in such games.

human experts. Although this was an important step in solving complex games, when complexity is increased even more (as for example by increasing the number of buckets), learning time and memory requirements will become impractical.

Another method by Hoda et. al. (2010) and Sandholm (2010) is to use the Excessive Gap Technique applied to a relaxed optimization problem from which the linear program described above is derived. The optimization problem is smoothened to be made differentiable; the solution to the new, relaxed problem is suboptimal in the original problem by some amount $\epsilon_0$. Parameters of the optimization problem are then modified by following the gradient of the smooth approximations of the objective functions. At iteration $i+1$ the modified parameters give a new solution with improved suboptimality $\epsilon_{i+1} < \epsilon_i$. This process is repeated until the desired value of $\epsilon$ is reached.

### 3.2 Sampling Algorithms for Computing Approximate Nash Equilibria

Performing Monte-Carlo sampling can enable algorithms to deal with highly complex domains. In this section we will discuss two algorithms that perform Monte-Carlo sampling, namely Monte-Carlo Tree Search (MCTS) and Monte-Carlo Counterfactual Regret Minimization (MCCFR). Although the internal workings of both techniques are different, they share the general procedure of sampling a simulated game, determining utilities at a leaf node (i.e., game state that ends the game), and backpropagating the results. The underlying idea of both techniques is to improve the quality of simulations progressively by taking into account the simulated games previously played. More specifically, simulations are driven to that part of the game tree that is most relevant, assuming players take rational decisions (i.e., choose actions that optimize their reward). The result of each game is backpropagated on the visited path. Progressively, the program concentrates its search on the best actions, leading to a deeper look-ahead ability.

We will now discuss an example in the game of Poker, as illustrated in Figure 2. As a first step (moving down), chance nodes and action nodes are sampled until a terminal node is reached. Chance nodes in Poker represent the dealing of cards. Similarly to previous work in Poker, we also apply bucketing on cards in our work. Cards are grouped together in buckets based on their strategic strength. We use a 10-bucket discretization (equal width), where EHS values (see Section 3.1) in the range of 0 to 0.1 are grouped together in bucket 1, 0.1 to 0.2 in bucket 2, and so forth. A higher bucket indicates a stronger hand. In our example we sampled buckets 5 and 7 in the preflop phase for respectively player one and two. Again, these buckets now reflect the strategic strength of players' private cards. With the appearance of new board cards in subsequent phases, we again encounter chance nodes, and buckets may change.[4]

We assume **imperfect recall**. With imperfect recall previous action or chance nodes are forgotten, and as a consequence several information sets are grouped together. For example, in our work we only take into account the current bucket assignment in information sets, and forget previous ones. This way we reduce the game complexity tremendously, which reduces both memory requirements and convergence time. In the case of imperfect recall the CFR algorithm loses its convergence guarantees to the NE, but even though we lose theoretical guarantees, this application of imperfect recall has been shown to be practical, specifically in Poker (Waugh, Zinkevich, Johanson, Kan, Schnizlein, & Bowling, 2009).

4. In the preflop phase the first player had the weakest hand, but with the appearance of board cards the player could, for example, have formed a pair and its bucket therefore has increased.

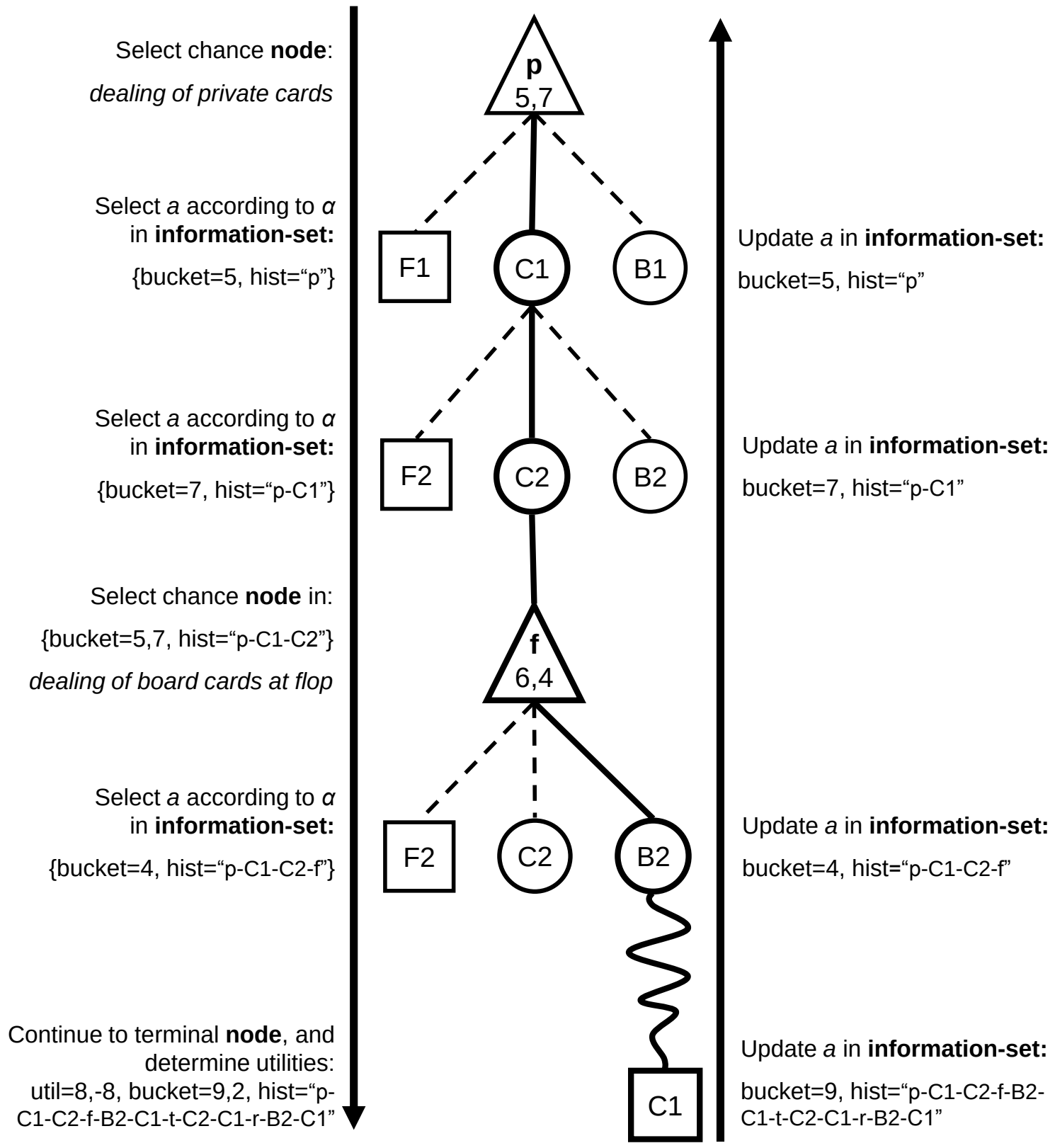

Figure 2: Illustration of Monte-Carlo simulations in the game of Poker. On the left going down, chancenodes (triangles) and action nodes (circles) are sampled based on some statistics at the information set level. Utilities are then determined at the terminal nodes (squares), having full information on both players' cards (i.e., buckets). On the right going up, the results of this simulated game are then backpropagated along all information sets that are a prefix of the terminal node. Statistics are updated at information set level, effectively altering the strategy.

Players select their actions $a$ based on the current strategy $\sigma$ given the information available to them (i.e., not knowing the bucket for the opponent player). The letters 'F', 'C' and 'B', followed by the player index, respectively correspond to the fold, call/check or bet/raise actions in the game of Poker. This process continues until a terminal leaf node is encountered, where utilities are determined for this simulated game.

As a second step (moving up), the utilities are backpropagated along the sampled path, and statistics are stored at the information set level. Different statistics are required for either MCCFR or MCTS. We discuss both algorithms in detail in the forthcoming subsections.

### 3.2.1 MONTE-CARLO TREE SEARCH

MCTS is a game tree search algorithm based on Monte-Carlo simulations. MCTS converges to a NE in perfect information games, whereas for the imperfect information case such guarantees are not given. It has been applied successfully in several perfect-information games (Lee et al., 2010; Chaslot et al., 2006; Bouzy & Chaslot, 2006; Coulom, 2006). It is therefore interesting to see if MCTS can also be successfully applied to the imperfect-information game of Poker. Two statistics of nodes are important for sampling actions in MCTS, i.e.:

1. the *value*, $v_a$ of action node $a$. This is the average of the reward of all simulated games that visited this node.
2. the *visit count*, $n_a$ of action node $a$. This represents the number of simulations in which this node was reached.

A crucial distinction with work that used MCTS in perfect-information games, is that now we assume *imperfect information*, as for example opponent cards in Poker. As a result, one has to reason over information sets (see Section 2.1) instead of individual nodes. Therefore, part of the algorithm is performed on information sets rather than individual nodes. The starting state of the game is represented by the root node, which is initially the only node in the tree. MCTS consists of repeating the following four steps (illustrated in Figure 3), as long as there is time left.

1. *Selection*. Actions in set $A$ are encoded as nodes in the game tree. They are chosen according to the stored statistics in a way that balances between exploitation and exploration. When exploiting, actions that lead to the highest expected value are selected. Less promising actions still have to be explored due to the uncertainty of the evaluation (exploration). We use the Upper Confidence Bound applied to Trees (UCT) rule to select actions (Kocsis & Szepesvári, 2006). In UCT, $a \in A$ is the set of nodes (possible actions) reachable from the parent node, $p$. Using the following equation, UCT selects the child node $a^*$ of parent node $p$ which has the highest value.

$$a^* \in \text{argmax}_{a \in A} \left( v_a + C \times \sqrt{\frac{\ln n_p}{n_a}} \right). \quad (2)$$

   Here $v_a$ is the expected value of the node $a$, $n_a$ is the visit count of $a$, and $n_p$ is the visit count of $p$, the parent of node $a$. $C$ is a coefficient that balances exploration and exploitation. A higher value encourages longer exploration since nodes that have not been visited often receive a higher value. This value is usually tweaked in preliminary experiments. Again, note that in imperfect-information games, expected values and visit counts are stored and updated per information set.

2. *Expansion*. When a leaf node is selected, one or several nodes are added to the tree. As such the tree grows with each simulated game. Please note that the tree in memory deals with game nodes (assuming full information), and not information sets (which are used in the selection and backpropagating part of the algorithm).

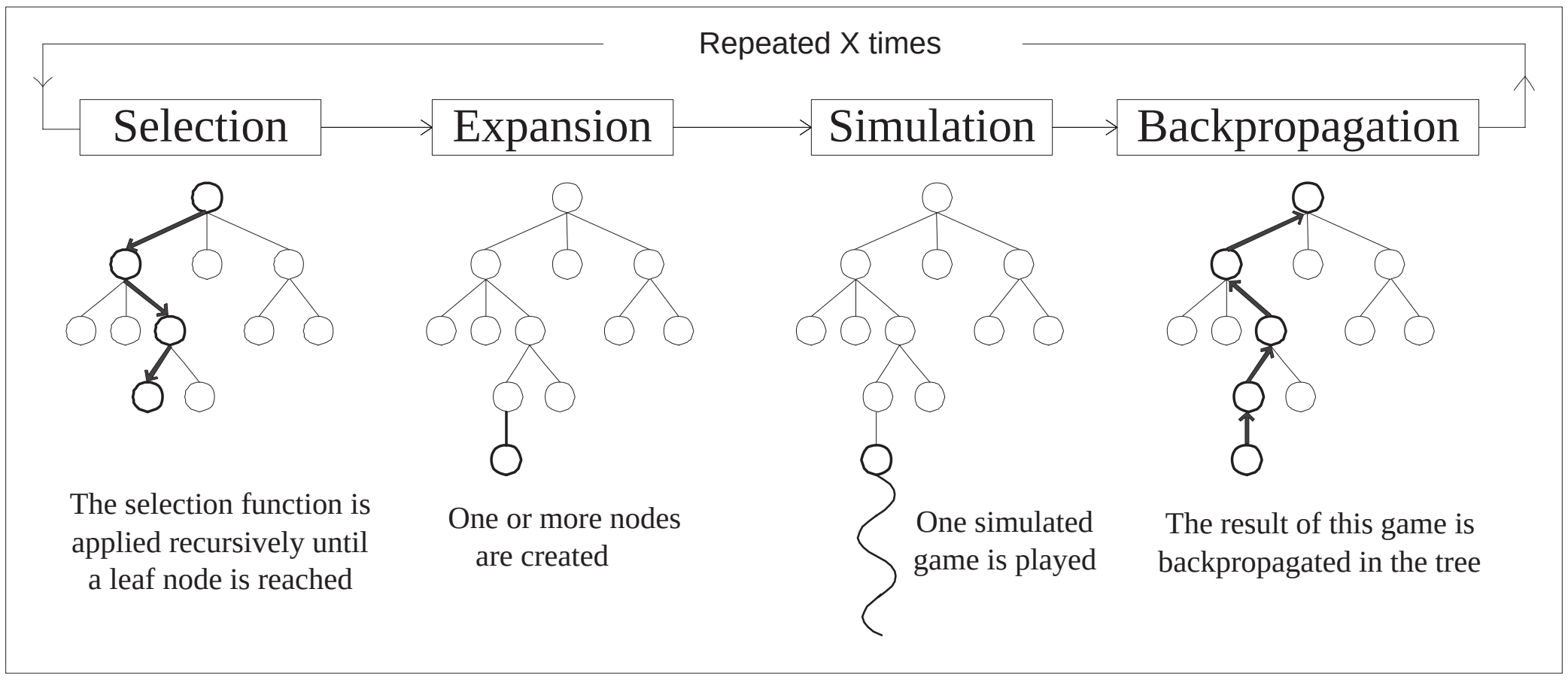

Figure 3: Outline of Monte-Carlo Tree Search (Chaslot et al., 2008).

3. *Simulation*. From the newly expanded node, nodes are selected according to some simulation policy until the end of the game. More realistic simulations will have a significant effect on the computed expected values. Examples for simulation strategies in Poker are: (1) *random simulations*, (2) *roll-out simulations* or (3) *on-policy simulations*. In the first simulation strategy, one samples nodes uniformly from the set of $A$. In roll-out simulations, one assumes that all remaining active players *call* or *check* until the end of the game. In other words, all remaining active players stay active and compete for the pot, but the stakes are not raised. Finally, the on-policy simulation uses current estimations of expected values or action probabilities to select actions. In this paper, we employ the latter. More specifically we always take action $a^*$ according to Equation 2. This produces simulations that are closest to the actual strategy.

4. *Backpropagation*. After reaching a terminal node $z$ (i.e., the end of the simulated game), we establish the reward for all players (having full information on the buckets for all players). Then we update each information set that contains a prefix of the terminal history. The visit counts are increased and the expected values are modified according to the rewards obtained at the terminal node.

After the MCTS algorithm is stopped, the (potentially mixed) action distribution is determined by the visit counts, i.e., the actions in information sets with a high visit count have a larger probability of being selected. In perfect-information games, those actions are selected with the highest expected value, which in the end leads to an optimal payoff. For imperfect-information games, such as Poker, potentially mixed strategy distributions are required for obtaining the optimal payoff (e.g., see the NES for Kuhn poker described in Section 2.2). Therefore, the ratio between the actions selected determines the probability distribution.

#### 3.2.2 MONTE-CARLO COUNTERFACTUAL REGRET MINIMIZATION

Before we describe the Monte-Carlo Counterfactual Regret Minimization (MCCFR) algorithm, we first describe the intuition behind the Counterfactual Regret Minimization (CFR) algorithm, since MCCFR is based on it.

CFR employs a full game-tree traversal in self-play, and updates player' strategies at information sets at each iteration. Strategy updates are based upon regret minimization. Imagine the situation where players are playing with strategy profile $\sigma$. Players may regret using their strategy $\sigma_i$ against $\sigma_{-i}$ to some extent. In particular, for some information set $I$ they may regret *not* taking a particular action $a$ instead of following $\sigma_i$. Let $\sigma_{I\to a}$ be a strategy identical to $\sigma$ except $a$ is taken at $I$. Let $Z_I$ be the subset of all terminal histories where a prefix of the history is in the set $I$; for $z \in Z_I$ let $z[I]$ be that prefix. The counterfactual value $v_i(\sigma, I)$ is defined as:

$$v_i(\sigma, I) = \sum_{z \in Z_I} \pi^{\sigma}_{-i}(z[I]) \pi^{\sigma}(z[I], z) u_i(z). \tag{3}$$

The algorithm applies a no-regret learning policy at each information set over these counterfactual values (Zinkevich et al., 2008). Each player starts with an initial strategy and accumulates a counterfactual regret for each action at each information set $r(I, a) = v(\sigma_{I\to a}, I) - v(\sigma, I)$ through self-play. Minimizing the regret of playing $\sigma_i$ at each information set also minimizes the overall external regret, and so the average strategies approach a NE.

MCCFR (Lanctot et al., 2009) avoids traversing the entire game tree on each iteration while still having the immediate counterfactual regrets be unchanged *in expectation*. Let $\mathcal{Q} = \{Q_1, \ldots, Q_r\}$ be a set of subsets of $Z$, such that their union spans the set $Z$. These $Q_j$ are referred to as blocks of terminal histories. MCCFR samples one of these blocks and only considers the terminal histories in the sampled block. Let $q_j > 0$ be the probability of considering block $Q_j$ for the current iteration (where $\sum_{j=1}^{r} q_j = 1$). Let $q(z) = \sum_{j: z \in Q_j} q_j$, i.e., $q(z)$ is the probability of considering terminal history $z$ on the current iteration. The sampled counterfactual value when updating block $j$ is:

$$\tilde{v}_i(\sigma, I|j) = \sum_{z \in Q_j \cap Z_I} \frac{1}{q(z)} \pi^{\sigma}_{-i}(z[I]) \pi^{\sigma}(z[I], z) u_i(z) \tag{4}$$

Selecting a set $\mathcal{Q}$ along with the sampling probabilities defines a complete sample-based CFR algorithm. For example, the algorithm that uses blocks that are composed of all terminal histories whose chance node outcomes are equal is called chance-sampled CFR. Rather than doing full game-tree traversals the algorithm samples one of these blocks, and then examines only the terminal histories in that block.

Sampled counterfactual value matches counterfactual value in expectation (Lanctot et al., 2009). That is, $\mathbb{E}_{j\sim q_j}\left[\tilde{v}_i(\sigma, I|j)\right] = v_i(\sigma, I)$. So, MCCFR samples a block and for each information set that contains a prefix of a terminal history in the block, it computes the *sampled counterfactual regrets* of each action, $\tilde{r}(I, a) = \tilde{v}_i(\sigma^t_{(I\to a)}, I) - \tilde{v}_i(\sigma^t, I)$. These sampled counterfactual regrets are accumulated, and the player's strategy on the next iteration is determined by applying the regret-matching rule to the accumulated regrets (Hart & Mas-Colell, 2000). This rule assigns a probability to an action in an information set. Define $r^+_I[a] = \max\{0, r_I[a]\}$. Then:

$$\sigma(I, a) = \begin{cases} 1/|A(I)| & \text{if } \forall a \in A(I) : r_I[a] \le 0 \\ \frac{r^+_I[a]}{\sum_{a \in A(I)} r^+_I[a]} & \text{if } r_I[a] > 0 \\ 0 & \text{otherwise.} \end{cases} \tag{5}$$

Here $r_I[a]$ is the cumulative sampled counterfactual regret of taking action $a$ at $I$. If there is at least one positive regret, each action with positive regret is assigned a probability that is normalized over all positive regrets and the actions with negative regret are assigned probability 0. If all the regrets are negative, then Equation 5 yields $\sigma(I,a) = 0\ \forall a$ in this information set. To repair this, the strategy is then reset to a default uniform random strategy.

There are different ways to sample parts of the game tree. Here we will focus on the most straightforward way, **outcome sampling**, which is described in Algorithm 1. In outcome-sampling $\mathcal{Q}$ is chosen so that each block contains a single terminal history, i.e., $\forall Q \in \mathcal{Q}, |Q| = 1$. On each iteration one terminal history is sampled and only updated at each information set along that history. The sampling probabilities, $\Pr(Q_j)$ must specify a distribution over terminal histories. We specify this distribution using a *sampling profile*, $\sigma'$, so that $\Pr(z)= \pi^{\sigma'}(z)$. Note that any choice of sampling policy will induce a particular distribution over the block probabilities $q(z)$. As long as $\sigma'_i(a|I) > \epsilon$, then there exists a $\delta > 0$ such that $q(z) > \delta$, thus ensuring Equation 4 is well-defined.

The algorithm works by sampling $z$ using policy $\sigma'$, storing $\pi^{\sigma'}(z)$. In particular, an $\epsilon$-greedy strategy is used to choose the successor history: with probability $\epsilon$ choose uniformly randomly and probability $1 - \epsilon$ choose based on the current strategy. The single history is then traversed forward (to compute each player's probability of playing to reach each prefix of the history, $\pi_i^{\sigma}(h)$) and backward (to compute each player's probability of playing the remaining actions of the history, $\pi_i^{\sigma}(h,z)$). During the backward traversal, the sampled counterfactual regrets at each visited information set are computed (and added to the total regret). Here,

$$\tilde{r}(I,a) = \begin{cases} w_I(\pi^{\sigma}(z[I]a,z) - \pi^{\sigma}(z[I],z)) & \text{if } z[I]a \sqsubseteq z \\ -w_I \pi^{\sigma}(z[I],z) & \text{otherwise} \end{cases}$$

$$\text{where } \; w_I = \frac{u_i(z)\pi_{-i}^{\sigma}(z[I])}{\pi^{\sigma'}(z)}. \tag{6}$$

The algorithm requires tables to be stored at each information set; each table has a number of entries equal to the number of actions that can be taken at that information set. Therefore, if we denote $|A_i|$ as the maximum number of actions available to player $i$ over all their information sets, then the space requirement for MCCFR is $O(|\mathcal{I}_1||A_1| + |\mathcal{I}_2||A_2|)$. The time required by MCCFR, using outcome sampling, depends on the regret bounds and the desired $\epsilon$. To reach a fixed $\epsilon$-NE, with probability $1 - p$ the number of iterations required is

$$O\left(\frac{2}{p\delta^2} \cdot \frac{|A|M^2}{\epsilon^2}\right)$$

where $\delta$ is the smallest probability of sampling a terminal history over all histories, $|A|$ is the maximum available actions over all information sets, and $|M|$ is a balance factor depending on the relative number of decisions taken by each player throughout the entire game with the property that $\sqrt{|\mathcal{I}|} \leq M \leq |\mathcal{I}|$. In contrast, the full CFR algorithm requires $O(|A|M^2/\epsilon^2)$ iterations; however, the iterations of MCCFR require only sampling a single history whereas an iteration of CFR requires traversing the entire game tree in the worst case. In practice, one benefit of outcome sampling is that information gained on previous iterations is quickly leveraged on successive iterations.

Algorithm 1 shows the pseudocode of the entire algorithm. We refer the interested reader to the work of Lanctot et al. (2009), providing a more in-depth discussion, including convergence proofs.

```
Data: root_node
Data: Sampling scheme ε_greedy
Data: Initialize information set markers: ∀I, c_I ← 0
Data: Initialize regret tables: ∀I, r_I[a] ← 0
Data: Initialize cumulative strategy tables: ∀I, s_I[a] ← 0
Data: Initialize initial strategy: σ(I, a) = 1/|A(I)|
for t = 1, 2, 3 ... do
    current_node ← root_node
    SELECT:
    while (current_node ≠ terminal) do
        P ← REGRET_MATCHING(r_I), ∀a ∈ I
        P* ← ε_GREEDY(P)
        current_node ← Select(current_node, P*)
    end
    current_node ← Parent(current_node)
    UPDATE:
    while (current_node ≠ root_node) do
        foreach a ∈ A[I] do
            r̃ = r(I, a) (sampled counterfactual regret)
            r_I[a] ← r_I[a] + r̃
            s_I[a] ← s_I[a] + (t − c_I) π_i^σ(z[I]) σ_i(I, a)
        end
        c_I ← t
        σ ← REGRET_MATCHING(r_I)
        current_node ← Parent(current_node)
    end
end
```

**Algorithm 1:** Outcome-sampling with Monte-Carlo Counter-Factual Regret Minimization

### 3.2.3 MONTE-CARLO COUNTERFACTUAL REGRET EXAMPLE

We now provide an example of Algorithm 1 on Kuhn Poker, as shown in Figure 1. The algorithm starts on the first iteration, during the selection phase. At the root node (a chance node), the chance outcome K|Q is sampled with probability $\frac{1}{6}$. Following this node, the algorithm loads the information set belonging to the first player where they have received a King and no actions have been taken; let us call this information set $I_1$. Since no regret has been collected for actions at $I_1$ yet, $P$ is set to the uniform distribution $U = (\frac{1}{2}, \frac{1}{2})$, which represent probabilities for (pass, bet). Then, the sampling distribution is obtained $P^* = (1 - \epsilon)P + \epsilon U$. Note that regardless of the value of $\epsilon$, on this first iteration each action is equally likely to be sampled. An action is sampled from $P^*$; suppose it is a bet. Following this action, the algorithm loads the information set belonging to the second player where they have received a queen and the action made by the first player was a bet ($I_2$). Similarly, the algorithm constructs $P$ and $P^*$ as before – which have identical distributions as at $I_1$; suppose the pass action is sampled this time. Finally, this action is taken and the terminal node is reached. Therefore, the terminal history that was sampled on the first iteration, $z_1$, is the sequence: (K|Q, bet, pass).

In the update phase, the algorithm updates the information sets touched by the nodes that were traversed in the sample in reverse order. Note that the notation $z_1[I_1]$ represents the subsequence (K|Q), and $z_1[I_2]$ represents (K|Q, bet). The sampled counterfactual regret is computed for each action at $I_2$. Note that $u_2(z_1) = -u_1(z_1) = -1$. The opponents' reaching probability is

$$\pi^{\sigma}_{-i}(z_1[I_2]) = \frac{1}{6} \cdot \frac{1}{2} = \frac{1}{12}.$$

The probability of sampling $z_1$ is

$$\pi^{\sigma'}(z_1) = \frac{1}{6} \cdot \frac{1}{2} \cdot \frac{1}{2} = \frac{1}{24}.$$

Therefore, $w_{I_2} = -2$. The two remaining things to compute are the tail probabilities $\pi^{\sigma}(z_1[I_2], z) = \frac{1}{2}$ and for each $\pi^{\sigma}(z_1[I_2]a, z) = 1$, where $a$ is the pass action. Finally, from equation 6 we get

$$\tilde{r}(I_2, \text{pass}) = -2 \cdot (1 - \frac{1}{2}) = -1,$$

and

$$\tilde{r}(I_2, \text{bet}) = -(-2 \cdot \frac{1}{2}) = 1.$$

After the updates on line 13, the regret table $r_{I_2} = (-1, +1)$. The average strategy table is then updated. The reaching probability $\pi^{\sigma}_i(z_1[I_2])$ is the product of probabilities of the strategy choices of player 2, equal to 1 here since player 2 has not acted yet. The player's strategy is currently $\sigma_2(I_2, \text{pass}) = \sigma_2(I_2, \text{bet}) = \frac{1}{2}$. The current iteration $t = 1$ and $c_{I_2} = 0$, therefore after the average strategy updates the table $s_{I_2} = (\frac{1}{2}, \frac{1}{2})$. Finally $c_{I_2}$ is set to 1 and $\sigma(I_2) = (0, 1)$ according to equation 5.

The algorithm then proceeds to update tables kept at $I_1$, where

$$u_1(z_1) = 1, \pi^{\sigma}_{-i}(z_1[I_1]) = \frac{1}{6}, \text{ and } \pi^{\sigma'}(z_1) = \frac{1}{24}, \text{ therefore } w_{I_1} = 4.$$

The tail reaching probabilities are $\pi^{\sigma}(z_1[I_1], z) = \frac{1}{4}$, and $\pi^{\sigma}(z_1[I_1]a, z) = \frac{1}{2}$, where $a$ is the bet action. This leads to sampled counterfactual values of

$$\tilde{r}(I_1, \text{pass}) = -1, \text{ and } \tilde{r}(I_1, \text{bet}) = 1.$$

As before $c_{I_1}$ is incremented to 1. The average strategy table update is identical to the one that was applied at $I_2$. Since there are no previous actions taken by the player, the reaching probability is again 1, and the player's current strategy was uniform, as at $I_2$. Finally $t$ is incremented to 2 and the entire process restarts.

### 3.3 Experiments

In the current section, we will empirically evaluate the chosen algorithms in the smaller domain of Kuhn Poker and in full-scale Poker. We will present experimental results in Kuhn Poker in Section 3.3.1. Then we will give results in Poker in Section 3.3.2. Note that Kuhn and full-scale Poker are explained in Section 2.2.

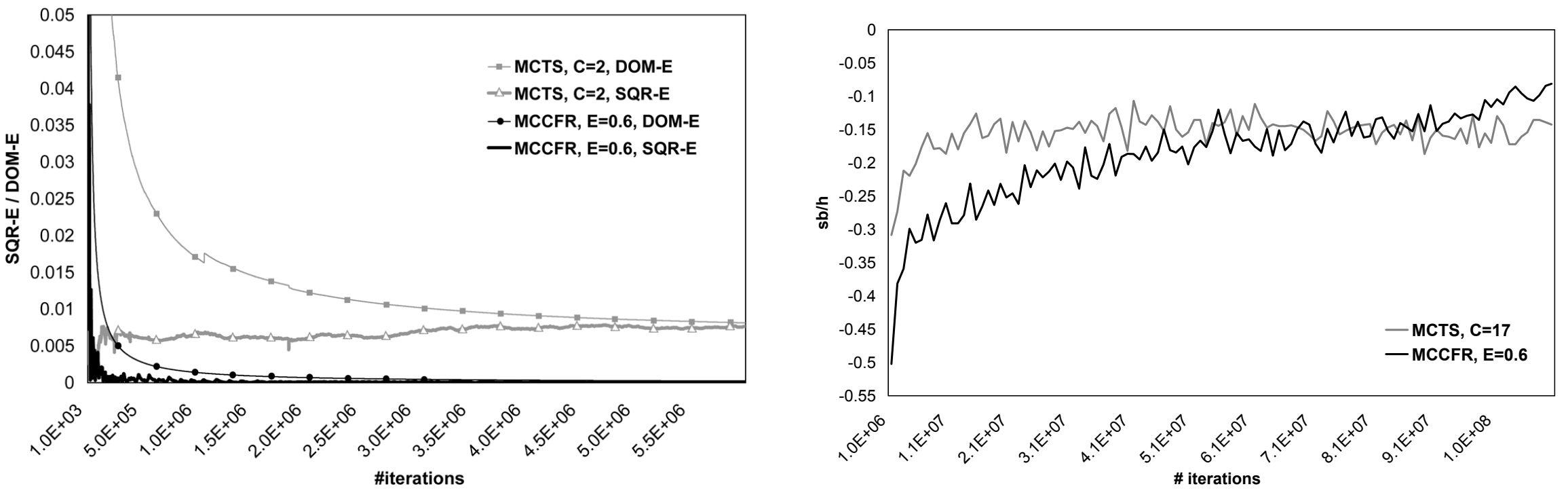

Figure 4: Experimental results for Monte-Carlo Counter-Factual Regret Minimization (MCCFR) and Monte-Carlo Tree Search (MCTS) in the game of Kuhn Poker (left) and Poker (right). In both figures the x-axis denotes the number of iterations the algorithms ran. The y-axis denotes the quality of the Nash-Equilibrium strategy learned so far. For our experiments in Kuhn Poker this is represented by the squared error (SQR-E) and cumulative dominated error (DOM-E), while in Poker we use the metric named small bets per hand (sb/h). For all metrics applies that a value close to zero indicates a near Nash-Equilibrium strategy. The opponent in the Poker experiments is an approximated Nash Equilibrium strategy, computed by MCCFR.

### 3.3.1 Kuhn Poker

We ran both MCCFR and MCTS in the game of Kuhn Poker. The $C$-constant for MCTS was tweaked in preliminary experiments and set to 2. It has been suggested by Balla & Fern (2009) that the $C$-constant should be set in the same scale as the payoff range. Also, Auer, Cesa-Bianchi, & Fischer (2002) discuss a modified version of the original Upper Confidence Bound (UCB) algorithm (on which the Upper Confidence Bound applied to Trees (UCT) algorithm is based) that tunes the exploration term. In our experiments, we ran several runs of MCTS with varying values for the parameter $C$, and then took the best experimental run. For MCCFR we used epsilon-greedy as its sampling scheme. As suggested in earlier work (Lanctot et al., 2009), $\epsilon$ was set to the relatively high value of $0.6$ to cover a large area of the search space.

After every $10^4$ iterations we measured the performance of the current policy. Since the equilibria are known in Kuhn Poker (see Section 2.2) we can compare our strategy with the theoretically correct one. Each evaluation we compute the *squared error*, which is simply the correct NE probability minus our current learned probability squared. We also compute the *cumulative dominated error*, which denotes the summed probabilities for selecting dominated actions.

From Figure 4 (left) we can confirm earlier results from the work of Lanctot et al. (2009) and Sturtevant (2008), namely that MCCFR learns a NE and MCTS learns a balanced situation that is not (necessarily) a NE. The balanced situation that MCTS eventually resides in depends on the parameter value $C$. Where MCCFR obtains squared errors and dominated errors close to zero, MCTS has converged slightly off the NE and is playing dominated actions. However, we also see that MCTS is still unlearning these dominated actions. Therefore, unlike MCCFR, MCTS does not learn perfectly rational strategies, and as such is exploitable. However, it does a reasonable job at avoiding dominated mistakes. As a consequence, it will not lose (much) to a NES.

#### 3.3.2 LARGER TEST DOMAIN: POKER

We evaluated policies learned by MCTS and MCCFR in the game of Poker. To reduce the complexity of the task of finding a NES, we decreased the size of the game-tree by applying a bucket discretization (Billings, 2006) on the cards, along with imperfect recall (i.e., buckets of previous phases are forgotten). At each phase, the strategic strength of private cards, along with zero or more board cards, determines the bucket.

**Learning Against a Precomputed Nash Equilibrium**
In this experiment we use a 10-bucket discretization for MCCFR and MCTS. MCCFR uses similar parameter settings compared to the experiments in Kuhn Poker, but for MCTS we change the $C$-constant to 17. Again, this value was determined in preliminary experiments. At every $10^6$ iterations we evaluate the current policy learned by MCTS and MCCFR. Policies are evaluated in small bets per hand ($sb/h$), which describes the big blinds won per hand on average; this quantity can thus be used to reflect a player's playing skill. The small bets per hand are computed in $10^5$ games against a pre-learned Nash-Equilibrium strategy (using MCCFR with a 10-bucket discretization). The results are shown in Figure 4 (right). The x-axis denotes the number of iterations, and the y-axis the small bets per hand. A value close to zero indicates that the strategy is no longer being exploited by the pre-computed NES, and arguably has converged to an (approximated) Nash Equilibrium itself.

We can see that MCTS again converges fast to a balanced situation, though not necessarily to the NE. MCCFR on the other hand converges considerably slower, but unlike MCTS it continues to learn better NE approximations. Given these results, one may conclude that MCTS learns reasonably good strategies fast, but MCCFR will in the long run produce better NE approximations.

It is interesting to mention here that a single iteration of MCTS requires much less computation time and memory than a single iteration of MCCFR. If the graph in Figure 4 (right) plotted computation time against performance, rather than the number of iterations required, we would see a larger advantage for MCTS in the early phase of learning. Clearly, the end result would still be the same; at a certain point, MCCFR surpasses MCTS.

**Playing Against Benchmark Poker Bots**
To get a good measure of the strength of the policies, we evaluated both policies against strong opponent bots provided with the software tool Poker Academy Pro, namely POKI and SPARBOT. For a more detailed explanation of POKI and experiments with this bot, we refer the reader to the work of Billings (2006). SPARBOT is a bot that plays according to the NE strategy described by Billings et al. (2003). It was designed solely for 2-player Poker, in contrast to POKI, which was designed for multi-player games. Since SPARBOT specializes in two-player games, it is significantly less exploitable in a two-player game than POKI, a stochastic rule-based bot.

We ran a large number of offline iterations for MCCFR and MCTS, froze the policies, and evaluated them. In addition to a 10-bucket discretization for MCTS and MCCFR, we also used MCCFR with a 100-bucket discretization to examine the effect of a finer abstraction. For clarity, we will mention the number of buckets used after the algorithm name; e.g., MCCFR100 refers to MCCFR with 100 buckets. All results are shown in Table 1. We performed $10^4$ evaluation games in the software tool Poker Academy Pro for each policy, with an estimated standard deviation of $0.06sb/h$. MCCFR10, after a great deal of iterations, wins by a small margin from POKI, while losing a small amount from SPARBOT. This loss may be due to our chosen abstractions (10-bucket imperfect recall), or due to our choice of using outcome sampling (Lanctot et al., 2009). Looking at MCCFR100, we see that the abstraction level is the culprit; it wins convincingly from POKI and SPARBOT. MCTS10 surprisingly, knowing it doesn't necessarily learn a NES, performs slightly better than MCCFR10 against POKI, although the difference is not significant. Against SPARBOT, the MCTS10 strategy loses slightly more than MCCFR10, with no statistical significance.

| Opponent | MCTS10 | MCCFR10 | MCCFR100 |
|---|---|---|---|
| POKI | 0.077 | 0.059 | 0.191 |
| SPARBOT | -0.103 | -0.091 | 0.046 |

Table 1: Experimental results for Monte-Carlo Counter-Factual Regret Minimization (MCCFR) and Monte-Carlo Tree Search (MCTS) against two Poker bots. Outcomes are reported in small bets per hand (sb/h). The numbers behind the algorithm names refer to the number of buckets used in the card abstraction.

To conclude, we evaluated two state-of-the-art algorithms in extensive-form games. We confirm previous theoretical claims with results obtained in experiments in a smaller game (Kuhn Poker), and for the first time apply both sampling techniques in the complex game of Poker. MCTS is a valid approach if one wants to learn reasonably strong policies fast, but these are not necessarily a NE, whereas MCCFR is the better choice if the quality of the strategy is more important than the learning time and memory constraints. However, both techniques can produce strategies that are competitive with strong Poker bots, especially if fine abstractions are used. Both sampling techniques show promise for complex domains, not exclusively Poker.

## 4. Robust Best-Response Learning via Monte-Carlo Restricted Nash Response

It is well-known that a perfectly rational Nash-Equilibrium strategy (NES) is not necessarily a best-response strategy (i.e., best counter-strategy) against strategies other than the rational strategy itself (Osborne & Rubinstein, 1994). If the opposition employs clearly inferior strategies, these can be exploited best using tailored counter-strategies. The resulting best-response strategy would be more profitable than the Nash-Equilibrium strategy (NES), because it is designed to win instead of designed not to lose. This stresses the importance of opponent modeling in games too complex to fully analyze, since we can expect that our opposition is incapable of playing a perfectly rational strategy, and we can profit from playing best-response strategies. However, we do not want to become too exploitable ourselves by other strategies, because our opponent model may be inaccurate, or players may switch strategies.

This section presents Monte-Carlo Restricted Nash Response (MCRNR), a sample-based algorithm for the offline computation of restricted Nash strategies in complex extensive-form games. A restricted Nash strategy is essentially a robust best-response strategy, i.e., it exploits opponents to some degree based on some opponent model, while preventing that the strategy itself becomes to exploitable. The new algorithm described in this section combines a state-of-the-art algorithm and a general technique, i.e., Monte-Carlo Counterfactual Regret Minimization (MCCFR) (Lanctot et al., 2009) (see Section 3.2.2) and Restricted Nash Response (RNR) (Johanson et al., 2008; Jo-hanson & Bowling, 2009). Given the promising results of applying sampling in Counterfactual Regret Minimization (CFR), we apply the original Restricted Nash Response (RNR) technique using Monte-Carlo Counterfactual Regret Minimization (MCCFR) as the underlying equilibrium solver. Our new algorithm, Monte-Carlo Restricted Nash Response (MCRNR), benefits from sampling only relevant parts of the game tree. The algorithm is therefore able to converge very quickly to robust best-response strategies given a model of the opponent(s).

The section is structured as follows. We first outline related work on computing best-response strategies. Then, we introduce our new algorithm, Monte-Carlo Restricted Nash Response (MCRNR). Finally, we describe experiments that validate the new algorithm in a variety of games, as discussed in Section 2.2.

## 4.1 Computing Best-Response Strategies

Poker is a perfect domain for investigating best-response strategies since the ability to anticipate an opponent's move highly influences the outcome of the game. We mention a few of the previous approaches to perform opponent modeling in Poker. Then, we discuss work on Restricted Nash Response (RNR).

### 4.1.1 General Opponent Modeling

One approach is the Adaptive Imperfect Information game-tree search algorithm (Billings, 2006), which has an opponent model integrated in it. It keeps track of statistics for both the outcome of the game and actions at opponent decision nodes for every possible betting sequence. The problem with this approach is that it uses little generalization and hence the frequency counts are limited to a small number of situations. A more general system for opponent modeling was obtained by training a neural network. Davidson, Billings, Schaeffer, and Szafron (2000) use nineteen different parameters as input nodes and three output nodes representing the possible actions. The input parameters include information about the players, information about the betting history and information on the community cards. Southey et al. (2005) use prior distributions over the opponent's strategy space and compute a posterior using Bayes' rule and observations of the opponent's decisions. It also investigates several ways to play an appropriate response to that distribution.

In this paper we also integrate an opponent model with a search technique in order to learn a best-response strategy. Unlike Billings (2006) we learn a model that generalizes between states. Our work also differs from the aforementioned studies in that we learn **robust** best-response strategies, which prevents the strategy from becoming too exploitable itself.

### 4.1.2 Restricted Nash Response

It is known that a best-response strategy, in other words a strategy that maximally exploits an opponent, is more profitable than a (pessimistic) NES, given that the model of the opponent is accurate. Experiments by Hoehn et al. in the game of Kuhn Poker, validate this claim. Johanson et al. (2008) argue that best-response strategies are not sufficiently robust; the best-response strategy may be exploited by strategies other than the strategy of the current opponent, or even by the opponent itself if the opponent model is not accurate. The authors therefore introduce a general technique named Restricted Nash Response (RNR) and put it to the test in Poker using chance-sampled Counterfactual Regret Minimization (CFR). Unlike Nash-Equilibrium strategies that are oblivious to opponent play, and best-response strategies that are potentially exploitable, RNR strategies are robust best-response strategies given a model of opponent policies (Johanson et al., 2008). It is shown that RNR strategies are capable of exploiting opponents, with reasonable performance even when the model is wrong. The RNR technique transforms an existing game into a modified game; an equilibrium solver then solves this modified game using the previously mentioned CFR algorithm, wherein it is assumed that the opponent plays according to a fixed strategy, as specified by the model, with a certain probability, denoted by the parameter $p$. Otherwise, the opponent plays according to a rational regret-minimizing strategy (see Figure 5).

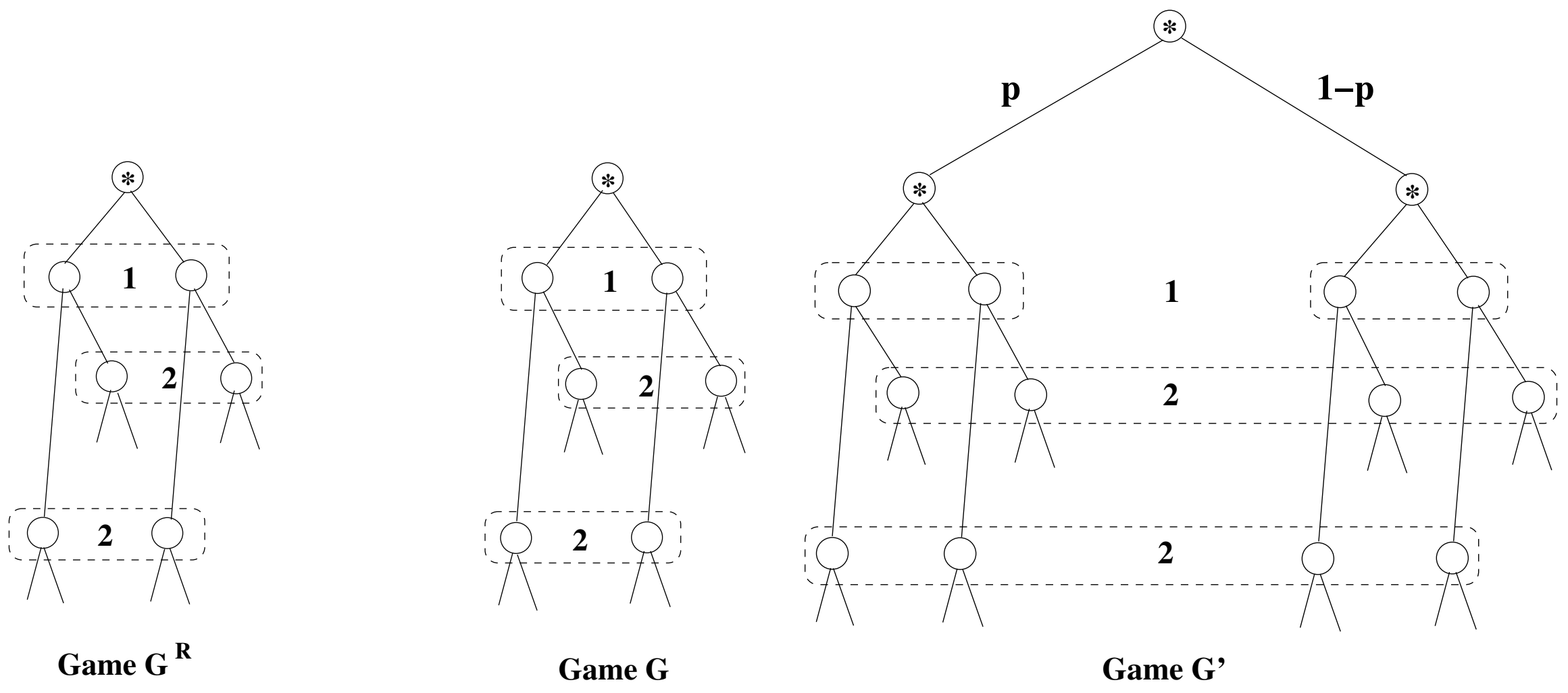

Figure 5: Illustration of Restricted Nash Response. The RNR technique transforms a game; a chance node with an outcome that is unrevealed (to the unrestricted player) is added to the top of both trees. This new game, $G'$, has two subtrees: the left subtree is $G^R$ and the right subtree is $G$. $G^R$ is identical to G except that one player is restricted to play the strategy $\sigma_{fix}$. Think of the initial coin flip as a moderator that either forces the restricted player to use $\sigma_{fix}$ or some other strategy $\sigma_i$. Their opponent does not know which of these two options was forced upon the restricted player. This causes the information sets containing the same nodes in each game to be merged, because the unrestricted player cannot tell them apart.

This $p$ parameter is an indication of how confident we are the model is correct; it can be thought of as a confidence value used to set the trade-off between **exploitation** and **exploitability**, where respectively the first indicates the maximum amount we can win from a specific opponent strategy, and the second indicates the maximum amount we risk losing by playing this strategy. Setting $p = 0$ leads to the unmodified game $G$ (see Figure 5) being solved, which results in a game-theoretic solution. The overall solution will not be exploitable (i.e., will at least break-even), but will also not maximally exploit its opponents. The other extreme, namely setting $p = 1$, will result in a pure best-response strategy against the fixed strategy denoted by the opponent model. In the game $G^R$, all opponent nodes have their action probabilities drawn from the opponent model. Focusing on this game results in a maximal exploitative strategy against the opponent model, potentially at the cost of becoming exploitable by other strategies.[5]

Calculating the RNR response requires a model of the opponent's strategy, denoted as $\sigma_{fix}$.[6] Suppose in a 2-player game, the opponent (i.e., restricted) player is player 2, then $\sigma_{fix} \in \Sigma_2$. Define $\Sigma_2^{p,\sigma_{fix}}$ to be the set of mixed strategies of the form $p\sigma_{fix} + (1-p)\sigma_2$ where $\sigma_2$ is an arbitrary strategy in $\Sigma_2$. The set of restricted best responses to $\sigma_1 \in \Sigma_1$ is:

$$BR^{p,\sigma_{fix}}(\sigma_1) = \operatorname*{argmax}_{\sigma_2 \in \Sigma_2^{p,\sigma_{fix}}} (u_2(\sigma_1, \sigma_2)) \tag{7}$$

A $(p, \sigma_{fix})$ RNR equilibrium is a pair of strategies $(\sigma_1^*, \sigma_2^*)$ where $\sigma_2^* \in BR^{p,\sigma_{fix}}(\sigma_1^*)$ and $\sigma_1^* \in BR^{p,\sigma_{fix}}(\sigma_2^*)$. In this pair, the strategy $\sigma_1^*$ is a $p$-restricted Nash response to $\sigma_{fix}$. These are counter-strategies for $\sigma_{fix}$, where $p$ provides a balance between exploitation and exploitability. In fact, given a particular value of $p$, solving a RNR-modified game assures the best possible trade-off between best response and equilibrium is achieved.

### 4.2 Monte-Carlo Restricted Nash Response (MCRNR)

We extend the original RNR algorithm with sampling. The resulting new algorithm, MCRNR, benefits from sampling only relevant parts of the game tree. It is therefore able to converge very fast to robust best-response strategies. The pseudo-code of the algorithm is provided in Algorithm 2. The algorithm has identical inputs compared to the MCCFR algorithm (see Algorithm 1), with an additional two components: (1) an opponent model that translates to a fixed strategy $\sigma_{fix}$ for the so-called restricted player $p_r$, and (2) a confidence value, $p$, that we assign to this model. For learning the opponent models, we apply a standard supervised learning method, namely the **J48** decision tree learner in the Weka datamining tool. Similar to the work in the original RNR article (Johanson et al., 2008), we here use a fixed value of $p$.[7]

We then sample a terminal history $h \in Z$, either selecting actions based on a provided opponent model, or based on the strategy obtained by regret-matching. The REGRET_MATCHING routine assigns a probability to an action in an information set (according to Equation 5). The $\epsilon$-greedy sampling routine $S(\sigma)$ samples action $a$ with probability

$$\epsilon \frac{1}{|A(I)|} + (1-\epsilon)\sigma_i(I, a). \tag{8}$$

After sampling the action, we recursively call the MCRNR routine given the extended history, namely the history of $h$ after applying $a$ (see line 13). Once at a terminal node $z$, utilities are determined and backpropagated through all $z[I] \sqsubset z$.

Regret and average strategy updates are applied when the algorithm returns from the recursive call, in lines 14 to 18. On line 15 we add the sampled counterfactual regret (according to Equation 6; it takes as input the reaching probabilities and utility for this sampled terminal history) to

5. Note the generality of the RNR technique: since it simply modifies an extensive-form game, the underlying solution technique used to solve the game is independent of the application of RNR. Nonetheless, from this point on we refer to RNR as an algorithm to mean the original application of the RNR technique coupled with the CFR algorithm.
6. We refer the reader back to Section 2.1 for an overview of the notations used in this section.
7. When learning the model from data it makes sense to have different values per information set because the confidence depends on how many observations are available per information set; counter-strategies using a model built from data are called Data-Biased Responses (Johanson & Bowling, 2009).

**initialize**: Information set markers: $\forall I, c_I \leftarrow 0$
**initialize**: Regret tables: $\forall I, r_I[a] \leftarrow 0$
**initialize**: Strategy tables: $\forall I, s_I[a] \leftarrow 0$
**initialize**: Initial strategy: $\sigma(I, a) = 1/|A(I)|$
**input** : A starting history $h$
**input** : A sampling scheme $\mathcal{S}(\sigma)$ (e.g., $\epsilon$-greedy)
**input** : An opponent model or fixed strategy $\sigma_{fix}$ for restricted player $p_r$
**input** : Confidence value $p$
**input** : Current iteration $t$
1 **MCRNR(h) =**
2 **if** $h \in Z$ **then**
3 | **return** $(u_i(h), \pi^\sigma(h))$
4 **else**
5 | $p_i \leftarrow P(h)$
6 | **if** *at chance node* **then** select chance node
7 | **else if** $p_i = p_r$ **then**
8 | | **if** *h is a prefix of a terminal history in the restricted subtree* **then** $\sigma_i(I) \leftarrow \sigma_{fix}(I)$
9 | | **else** $\sigma_i(I) \leftarrow$ REGRET_MATCHING$(r_{I_i})$
10 | **else**
11 | | $\sigma_i(I) \leftarrow$ REGRET_MATCHING$(r_{I_i})$
12 | Sample $a$ from $\mathcal{S}(\sigma_i(I))$
13 | $(u, \pi) \leftarrow$ **MCRNR(h+a)**
14 | **foreach** $a \in A(I)$ **do**
15 | | $r_I[a] \leftarrow r_I[a] + \tilde{r}(I, a)$
16 | | $s_I[a] \leftarrow s_I[a] + (t - c_I)\pi_i^\sigma \sigma_i(I, a)$
17 | **end**
18 | $c_I \leftarrow t$
19 | **return** $(u, \sigma)$
20

**Algorithm 2:** One iteration of the Monte-Carlo Restricted Nash Response algorithm.

the cumulative regret. On line 16 the average strategy is updated using *optimistic averaging*, which assumes that nothing has changed since the last visit at this information set.[8] Finally, the strategy tables are updated before a new iteration starts. In the next iteration, another player becomes the restricted player. This process is repeated a number of times until satisfactory; meanwhile, each player has been assigned to be the restricted player. At any iteration, the average strategy $\overline{\sigma}(I, a)$ can be obtained by normalizing $s_I$. When $p_r = 2$ then $\sigma_1^* = \overline{\sigma}_1$. When $p_r = 1$ then $\sigma_2^* = \overline{\sigma}_2$. Over time, $\sigma^* = (\sigma_1^*, \sigma_2^*)$ approaches an RNR equilibrium.[9]

8. For more information on update strategies, we refer to the work of Lanctot et al. (2009).
9. Note that the Data-Biased Response (DBR) variant (Johanson & Bowling, 2009) works in a slightly different way. Instead of having the selection of restricting the player or not at the root as a chance node, it is done before each information set and then hidden from the restricted player (Johanson & Bowling, 2009). The restricted player is forced to used a mixed strategy based on the confidence value for the current information set $pConf$. Lines 8 and 9

### 4.3 Experiments

We performed experiments in smaller games and in Poker. In the smaller games we performed two types of experiments; one to characterize the relationship between exploitation and exploitability and the other for evaluating convergence rates of sampling versus non-sampling algorithms. In Poker we evaluate the strength of learned strategies against benchmark opponent players.

#### 4.3.1 SMALLER GAMES

We ran two separate sets of experiments for the games OCP (we use a deck of size $N = 500$), Goofspiel, Bluff, and PAM. The first set of experiments aims to characterize the relationship between exploitation and exploitability for different values of $p$. The second set of experiments is a comparison of the convergence rates of RNR and MCRNR. In both cases perfect opponent models were taken from runs of MCCFR and $\epsilon$ was set to $0.6$. Results are shown in Figures 6 and 7.

Results from the first set of experiments may influence the choice of $p$. If exploitation is much more important than exploitability then a value above $0.9$ is suggested; on the other hand a noticeable boost in exploitation can be achieved for a small loss of exploitability for $0.5 \leq p \leq 0.8$. In every game except Bluff it seems that the region $p \in [0.97, 1]$ has high impact on to the magnitude of this trade-off. Results from Figure 7 confirm the performance benefit of sampling since MCRNR produces a better NE approximation in less time, especially in the early iterations. This is particularly important when attempting to learn online (i.e. during playing, rather than beforehand), when time might be limited.

#### 4.3.2 LARGER TEST DOMAIN: POKER

We evaluated the policies learned by MCRNR against two poker bots, namely POKI and SPARBOT. Both opponents were also used in experiments with a game theoretic (or rational) player in Section 3. The experimental settings as well as chosen abstractions are identical to those in the experiments in Section 3.3.2. We remind the reader that SPARBOT is a bot designed to play according to a NES in an abstracted game; it is therefore less exploitable than POKI.

Our MCRNR implementation in Poker restricts the player at each information set, and in essence represents a MCDBR algorithm (see Section 4.2). However, unlike the original DBR paper, in our experiments $pConf$ is a global constant and its value is not biased from data. The resulting algorithm therefore is a mix between the two algorithms. We will continue using the name MCRNR here, but the reader should note that the original RNR algorithm is slightly different.[10]

For the opponent modeling, we deliberately chose a setup that was 'realistically difficult'. We observed only $\sim 20K$ games played against both POKI and SPARBOT, as opposed to, e.g., the 1 million games used by Johanson and Bowling (2009) for RNR. These games were used to gather opponent data concerning the two bots. Learning an opponent model can be approached as a pattern recognition task (Bishop, 2006), wherein a model is learned based on experience (in this case, previous Poker games of the players that are modeled). The model is then used to estimate the behavior of opponents in unseen situations. Since the opponent data we gathered is rather sparse due to the $20K$ games, and since a frequency count cannot generalize, we chose to apply a standard from Algorithm 2 are replaced by $\sigma_i(I) \leftarrow pConf \sigma_{fix}(I) + (1 - pConf)\sigma_i(I)$, where $pConf$ is a specific value of $p$ per information set.

10. Based on experimental results in the smaller games, we noticed that the difference between vanilla RNR and MCDBR with a fixed $pConf$ is very small – both algorithms are thus similar.

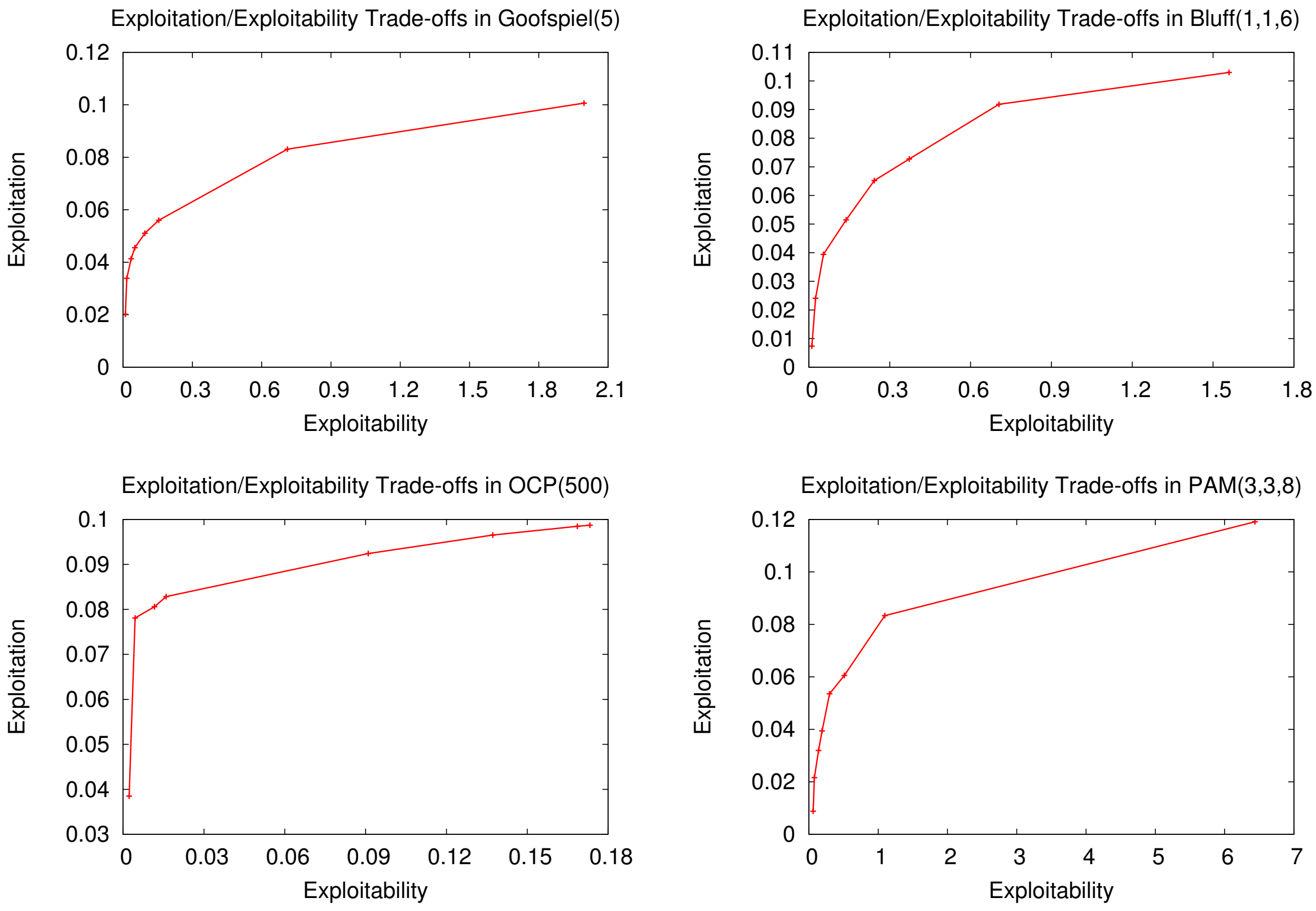

Figure 6: The trade-off between exploitation and exploitability with Monte-Carlo Restricted Nash Response. The exploitation value is the gain in payoff when using the Monte-Carlo Restricted Nash Response equilibrium profile compared to a Nash equilibrium profile, summed over $p_r \in \{1, 2\}$. The exploitability is $b_i(\sigma_{-i})$ summed over $i = p_r \in \{1, 2\}$. The value of $p$ used, from bottom-left point to top-right point, was: $0, 0.5, 0.7, 0.8, 0.9, 0.93, 0.97, 1$.

decision tree induction algorithm (i.e., the J48 algorithm in the Weka data-mining tool) to learn an opponent model from the sparse data. We provided the J48 algorithm with five simple features, namely (1) the starting seat relative to the button, (2) the sum of bets or raises during the game, (3) the sum of bets or raises in the current phase, (4) the sum of bets or raises of the modeled player in the game, and finally (5) the bucket of the modeled player (if it was observed). For each specific phase, we learn a model that predicts the strategy of the modeled player. We set $p$ to a fixed value of $0.75$.[11] We ran offline iterations of MCRNR, froze the policy, and evaluated it. All results are shown in Table 2, including the results for MCCFR, which we already presented earlier. We performed $10^4$ evaluation games for each player. Again, we provide mostly results with a 10-bucket abstraction (labeled 'MCRNR10'). A 100-bucket abstraction ('MCRNR100') was used as well, against POKI, to demonstrate the effect of finer abstraction levels.

11. This value is not adapted based on the experience in a specific information set, as was done in the data-biased approach (Johanson & Bowling, 2009).

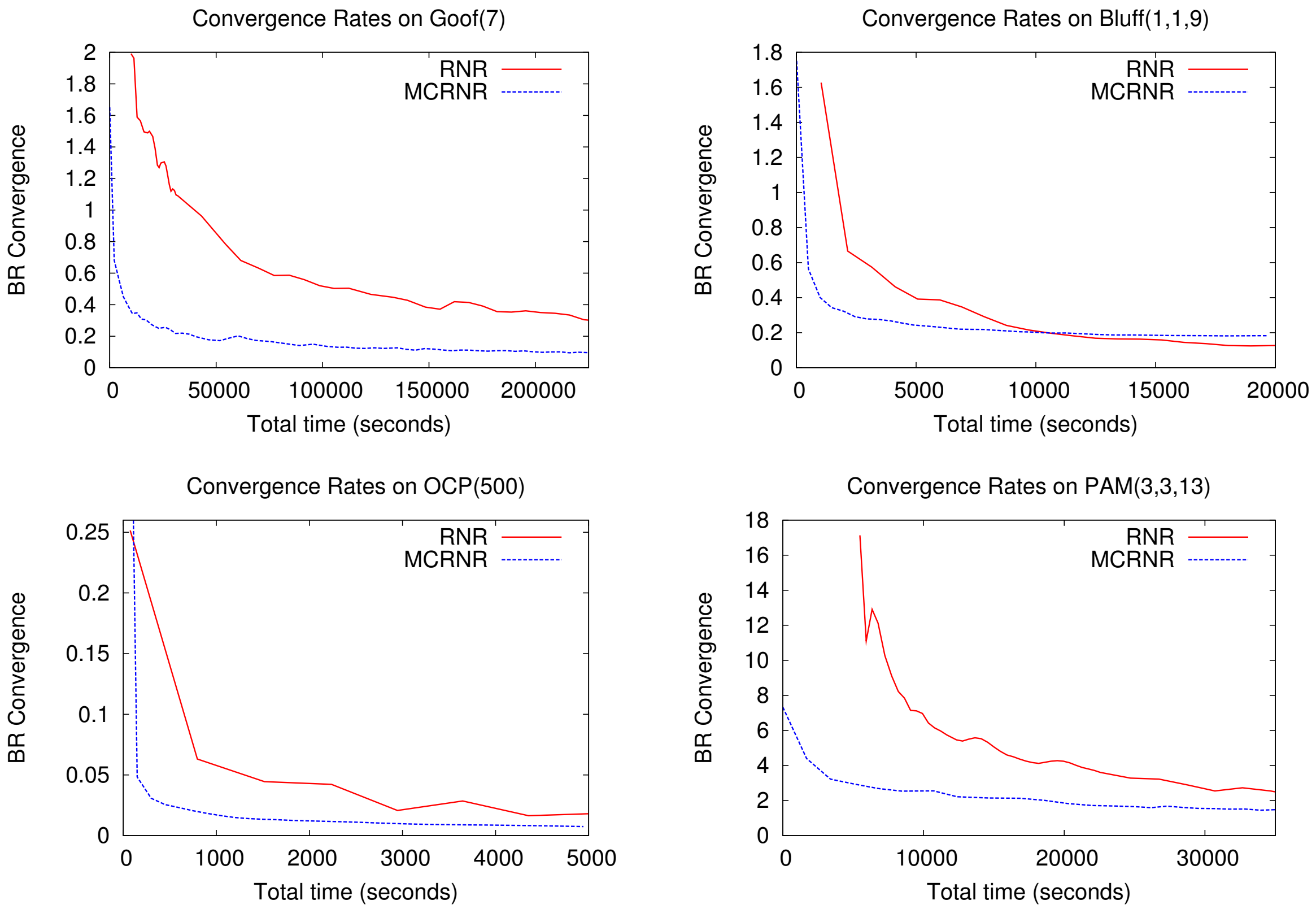

Figure 7: The convergence rates of Restricted Nash Response versus Monte-Carlo Restricted Nash Response. Fixed strategy profiles $\sigma_{fix}$ were generated using MCCFR and run until $\epsilon_{\sigma_{fix}} \leq 0.1$. Each data point in the graphs represent two separate runs with average profiles $(\overline{\sigma}_1^r, \overline{\sigma}_2)$ and $(\overline{\sigma}_1, \overline{\sigma}_2^r)$, where a superscript of $r$ represents the restricted player. The profile of interest is then $\overline{\sigma} = (\overline{\sigma}_1, \overline{\sigma}_2)$. The value on the y-axis is $\epsilon_{\overline{\sigma}} = b_1(\overline{\sigma}_2) + b_2(\overline{\sigma}_1)$. The profile $\overline{\sigma}$ is an RNR equilibrium when $\epsilon_{\overline{\sigma}}$ is minimized. Note that $\epsilon_{\overline{\sigma}}$ does not necessary approach 0 when $p > 0$; the strategies may be always be somewhat exploitable due to opponent exploitation.

As expected, MCRNR exploits POKI considerably more than MCCFR, namely with $0.369\ sb/h$ (using a 10-bucket abstraction) and $0.482\ sb/h$ (100 buckets). Interestingly, as depicted in Figure 8, even MCRNR10 has learned to exploit POKI in only ~20 million sampled iterations. We note that in a sampled iteration, only very few nodes are touched (i.e., only information sets are updated along the history of the sampled terminal node), while in a RNR (and CFR, for that matter) iteration all information sets are updated. Consequently, 20 million sampled iterations with MCRNR (or MCCFR) map to far less full-backup iterations with RNR (or CFR), and require much less computation time than 20 million full backups.

Against SPARBOT, which plays a better NES approximation, an improvement in performance is also observed, but here, the difference is not significant. This is not surprising, given that SPAR-

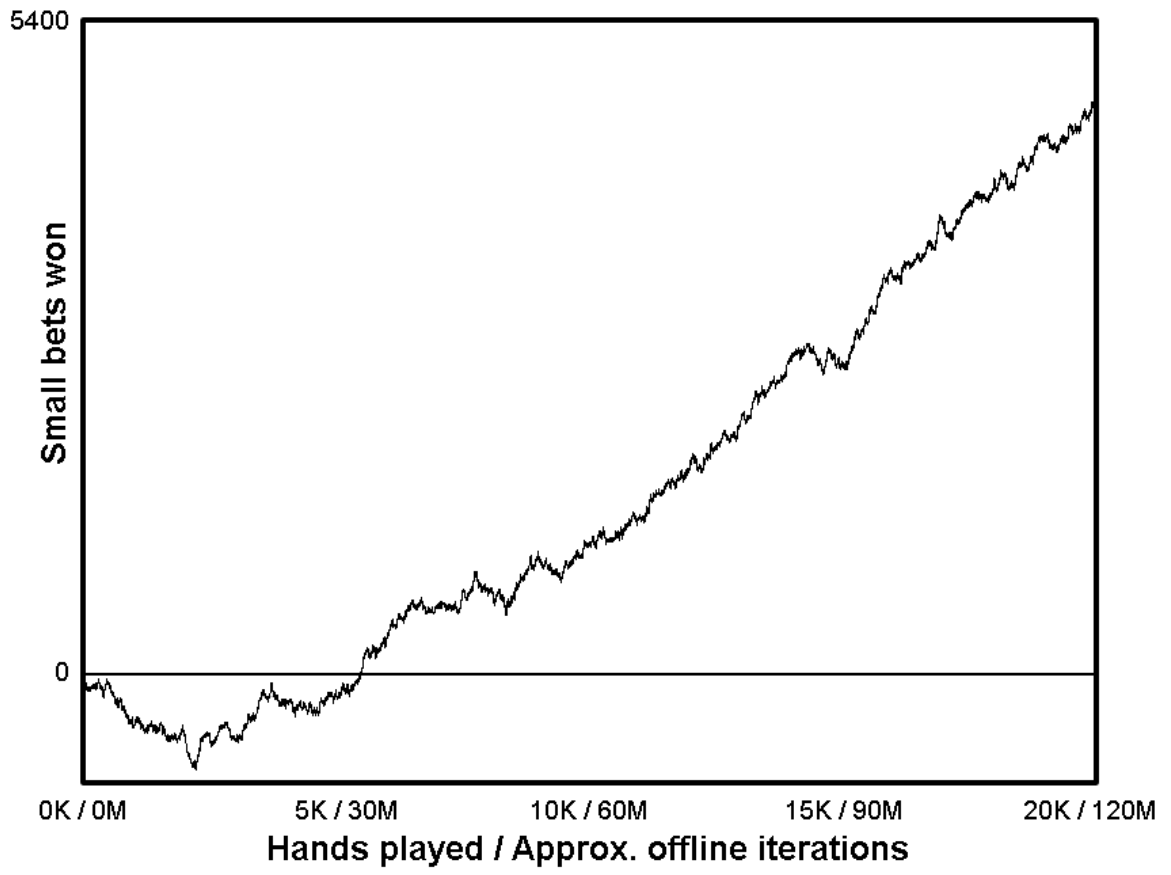

Figure 8: An online evaluation of a MCRNR policy during learning against POKI. While the bot is playing $1,000$ online games, an approximated 6 million offline iterations with Monte-Carlo Restricted Nash Response, using a 10-bucket abstraction, are run.

| Opponent | MCCFR10 | MCRNR10 | MCCFR100 | MCRNR100 |
|---|---|---|---|---|
| POKI | 0.059 | 0.369 | 0.191 | 0.482 |
| SPARBOT | -0.091 | -0.039 | 0.046 | 0.061 |

Table 2: Experimental results with MCRNR versus two bots. We repeat the results of Monte-Carlo Counter-Factual Regret Minimization (MCCFR) as reported in Section 3. Outcomes are in small bets per hand (sb/h). Numbers behind algorithm names refer to the abstraction level (number of buckets).

BOT is not as (extremely) exploitable as Poki. Clearly, though, MCRNR performs similarly against SPARBOT as MCCFR, implying that the MCRNR policy plays very few dominated actions.

In conclusion, against both POKI as well as SPARBOT, MCRNR finds at least an equally good policy as MCCFR. Against POKI, an exploitable opponent, the benefit of the algorithm becomes apparent; the MCRNR policy earns much more than the MCCFR policy, even with a coarse abstraction (compare $0.191\, sb/h$ for MCCFR100 with $0.369\, sb/h$ for MCRNR10; a coarser abstraction).

We also deliberately limited the amount of data available for the opponent model to illustrate the performance of the new algorithm under difficult conditions. Under more ideal conditions, i.e., most prominently with more opponent data, the algorithm can be expected to further outperform MCCFR in terms of (simulated) money won when playing against an exploitable opponent, and to further outperform both RNR and MCCFR in terms of the time required to find a good policy.

## 5. Conclusion

This article highlights two of our contributions in the field of decision-making in complex partially observable stochastic games. We first applied two existing recent search techniques that use Monte-Carlo sampling to the task of approximating a Nash-Equilibrium strategy (NES), namely Monte-Carlo Tree Search (MCTS) and Monte-Carlo Counterfactual Regret Minimization (MCCFR). These algorithms have not been compared before in a game as complex as we use, namely two-player Limit Texas Hold'Em Poker. MCTS has been used predominantely in perfect-information games such as Go. In such games, the algorithm is proven to compute a NES. In imperfect-information games, balanced situations are learned that are not necessarily a Nash-Equilibrium (NE) (Sturtevant, 2008). In our experiments, we confirm this finding: the algorithm can indeed learn reasonably strong policies in Poker, with the drawback that these strategies are not necessarily a NE. MCCFR on the other hand has already been shown to converge to a NE in smaller imperfect information games, such as the ones we outlined in Section 2.2 (Lanctot et al., 2009). We apply it for the first time in the complex game of Poker, and show that it indeed does approximate a NES. The initial convergence of MCCFR is slower than that of MCTS. This may be due to the fact that we used outcome sampling. Other sampling schemes, e.g., external sampling (Lanctot et al., 2009), may lead to MCCFR converging as fast as MCTS (where 'fast' is measured in number of iterations required). It should be mentioned that a typical iteration of MCCFR takes significantly longer (in actual time) than one of MCTS, due to additional computational complexity involved in the backpropagation process. MCCFR also takes up more memory because more statistics are required for computing strategy distributions.

Our second contribution relates to the observation that Nash-Equilibrium strategies are not necessarily best to deal with clearly irrational opposition (i.e., players not playing a NES). A tailored best-response strategy can yield more profit. Pure best-response strategies however may be brittle and exploitable against strategies other than the one they were trained against. We present Monte-Carlo Restricted Nash Response (MCRNR), a sample-based algorithm for the computation of restricted Nash strategies, essentially robust best-response strategies that (1) exploit irrational opponents more compared to the NES and (2) are not (too) exploitable by other strategies. This algorithm combines the advantages of two state-of-the-art existing algorithms, i.e., MCCFR and Restricted Nash Response (RNR). MCRNR samples only relevant parts of the game tree. It is therefore able to converge faster to robust best-response strategies than RNR. We evaluate our algorithm on a variety of imperfect-information games that are small enough to solve yet large enough to be strategically interesting. We empirically show that MCRNR learns much quicker than standard RNR in smaller games. We also apply MCRNR in the large game of Poker, deliberately choosing hard settings, i.e., relatively few iterations to come up with a policy, and relatively little opponent data. Even with such hard settings, MCRNR learns to exploit the strong (yet exploitable) opponent bot POKI significantly more than the NES learned by MCCFR, while performing similarly as MCCFR against the strong (hardly exploitable) opponent bot SPARBOT. MCRNR achieves this performance in a fraction of the computation time required by previous algoritms.

The strong results obtained by MCCFR and MCRNR in the complex game of two-player Poker point to many possible applications for (refinements of) these algorithms. Most prominently, we see ample opportunity for continued work in Poker. One of the first applications of interest is two-player Poker with even finer abstractions. We show that a 100-bucket abstraction performs much better than a 10-bucket one. Indeed, state-of-the-art two-player NE bots use strategies computed with extremely fine-grained abstractions. Less abstractions would further improve performance considerably against the opponent bots used in this article. However, in such settings, we would require more opponent-model data, and also need to allow more iterations to learn policies. Second, integrating the Data-Biased Response (DBR) approach may lead to increased performance. Third, it would be highly interesting to evaluate the performance of the algorithms in Poker with more than two players. In zero-sum games with more than two players, the NES is no longer guaranteed not to lose; coalitions between players may be formed. Nonetheless, the first work on applying algorithms such as CFR to games with more than two players suggests that the NES still is a very strong strategy (Risk & Szafron, 2010). It would be very interesting to determine whether MCCFR and MCRNR are as beneficial (for the speed of convergence and the quality of the solution converged to) in three-player games as in a two-player game.

## Acknowledgments

The authors thank Michael Johanson for his extensive commentary on the article, and his valuable input on the Restricted Nash Response algorithm and Poker research at the University of Alberta in general. We also thank the anonymous reviewers and the editors for their valuable input.